\documentstyle[]{mn}
\newif\ifAMStwofonts

\ifoldfss
  \newcommand{\rmn}[1] {{\rm #1}}

  \ifCUPmtlplainloaded \else
    \NewTextAlphabet{textbfit} {cmbxti10} {}
    \NewTextAlphabet{textbfss} {cmssbx10} {}
    \NewMathAlphabet{mathbfit} {cmbxti10} {} 
    \NewMathAlphabet{mathbfss} {cmssbx10} {} 
  \fi
  \ifAMStwofonts
    \ifCUPmtlplainloaded \else
      \NewSymbolFont{upmath} {eurm10}
      \NewSymbolFont{AMSa} {msam10}
      \NewMathSymbol{\upi}     {0}{upmath}{19}
      \NewMathSymbol{\umu}     {0}{upmath}{16}
      \NewMathSymbol{\upartial}{0}{upmath}{40}
      \NewMathSymbol{\leqslant}{3}{AMSa}{36}
      \NewMathSymbol{\geqslant}{3}{AMSa}{3E}

    \fi
  \fi
\fi 

\ifnfssone
  \newmathalphabet{\mathit}
  \addtoversion{normal}{\mathit}{cmr}{m}{it}
  \addtoversion{bold}{\mathit}{cmr}{bx}{it}
  \newcommand{\rmn}[1] {\mathrm{#1}}

  \newmathalphabet{\mathbfit} 
  \addtoversion{normal}{\mathbfit}{cmr}{bx}{it}
  \addtoversion{bold}{\mathbfit}{cmr}{bx}{it}
  \newmathalphabet{\mathbfss} 
  \addtoversion{normal}{\mathbfss}{cmss}{bx}{n}
  \addtoversion{bold}{\mathbfss}{cmss}{bx}{n}
  \ifAMStwofonts
    \ifCUPmtlplainloaded \else
      \UseAMStwoboldmath
      \makeatletter
      \new@mathgroup\upmath@group
      \define@mathgroup\mv@normal\upmath@group{eur}{m}{n}
      \define@mathgroup\mv@bold\upmath@group{eur}{b}{n}
      \edef\UPM{\hexnumber\upmath@group}
      \new@mathgroup\amsa@group
      \define@mathgroup\mv@normal\amsa@group{msa}{m}{n}
      \define@mathgroup\mv@bold\amsa@group{msa}{m}{n}
      \edef\AMSa{\hexnumber\amsa@group}
      \makeatother
      \mathchardef\upi="0\UPM19
      \mathchardef\umu="0\UPM16
      \mathchardef\upartial="0\UPM40
      \mathchardef\leqslant="3\AMSa36
      \mathchardef\geqslant="3\AMSa3E
    \fi
  \fi
\fi 

\ifnfsstwo
  \newcommand{\rmn}[1] {\mathrm{#1}}

  \DeclareMathAlphabet{\mathbfit}{OT1}{cmr}{bx}{it}
  \SetMathAlphabet\mathbfit{bold}{OT1}{cmr}{bx}{it}
  \DeclareMathAlphabet{\mathbfss}{OT1}{cmss}{bx}{n}
  \SetMathAlphabet\mathbfss{bold}{OT1}{cmss}{bx}{n}
  \ifAMStwofonts
    \ifCUPmtlplainloaded \else
      \DeclareSymbolFont{UPM}{U}{eur}{m}{n}
      \SetSymbolFont{UPM}{bold}{U}{eur}{b}{n}
      \DeclareSymbolFont{AMSa}{U}{msa}{m}{n}
      \DeclareMathSymbol{\upi}{0}{UPM}{"19}
      \DeclareMathSymbol{\umu}{0}{UPM}{"16}
      \DeclareMathSymbol{\upartial}{0}{UPM}{"40}
      \DeclareMathSymbol{\leqslant}{3}{AMSa}{"36}
      \DeclareMathSymbol{\geqslant}{3}{AMSa}{"3E}
    \fi
  \fi
\fi 

\ifCUPmtlplainloaded \else
  \ifAMStwofonts \else 
    \def\upi{\pi}
    \def\umu{\mu}
    \def\upartial{\partial}
  \fi
\fi

\newcommand{\grb}
	{GRB}
\newcommand{\grbs}
	{GRBs}

\newcommand{\dm}
	{dark matter}

\newcommand{\mond}
	{MOND}

\newcommand{\SDSS}
        {SDSS}

\newcommand{\samethanks}
	{{\Huge $^\star$}}

\newcommand{\vect}[1]
        {\mbox{\boldmath ${#1}$}}

\newcommand{\etal}
	{et al.}
\newcommand{\eg}
	{e.g.}
\newcommand{\cf}
	{cf.}
\newcommand{\ie}
	{i.e.}
\newcommand{\eq}[1]
	{equation~(\ref{equation:#1})}
\newcommand{\eqs}[1]
	{equations~(\ref{equation:#1})}

\newcommand{\sect}[1]
	{Section~\ref{section:#1}}

\newcommand{\fig}[1]
	{Fig.~\ref{figure:#1}}

\newlength{\singlefigureheight}
\newlength{\doublefigureheight}
\newlength{\triplefigureheight}
\newlength{\squarefigureheight}
\newcommand{\AaA}
        {A\&A}

\newcommand{\AJ}
        {AJ}
\newcommand{\ApJ}
        {ApJ}

\newcommand{\ARAA}
        {ARA\&A}

\newcommand{\MNRAS}
        {MNRAS}

\newcommand{\pointmass}
        {point-mass}

\newcommand{\pointmasses}
        {point-masses}

\newcommand{\los}
        {line-of-sight}

\newcommand{\LMC}
	{LMC}

\newcommand{\FL}
        {FL}

\newcommand{\gr}
	{GR}
\newcommand{\Gr}
	{GR}

\begin{document}

\title[Lensing constraints on gravity]
{Empirical constraints on alternative gravity theories from 
gravitational lensing}

\author[D.\ J.\ Mortlock and E.\ L.\ Turner]
       {
        Daniel J.\ Mortlock,$^{1,2}$\thanks{
		E-mail: mortlock@ast.cam.ac.uk (DJM);
		elt@astro.princeton. edu (ELT)}
	and Edwin L.\ Turner$^3$\samethanks\ \\
        $^1$Astrophysics Group, Cavendish Laboratory, Madingley Road,
        Cambridge CB3 0HE, U.K. \\
	$^2$Institute of Astronomy, Madingley Road, Cambridge
	CB3 0HA, U.K. \\
	$^3$Princeton University Observatory, Peyton Hall,
	Princeton, NJ 08544, U.S.A. \\
       }

\date{
Accepted. 
Received; in original form 2001 January 22} 

\pagerange{\pageref{firstpage}--\pageref{lastpage}}
\pubyear{2001}

\label{firstpage}

\maketitle

\begin{abstract}
If it is hypothesised that there is no \dm\ then 
some alternative gravitational theory must take the place of 
general relativity (\gr) on the largest scales.
Dynamical measurements can be used to investigate the nature
of such a theory, but only where there is visible matter.
Gravitational lensing is potentially a more powerful probe
as it can be used to measure deflections 
far from the deflector, and, for sufficiently large separations,
allow it to be treated as a \pointmass.
Microlensing within the local group does not yet provide
any interesting constraints, as only images formed close to
the deflectors are appreciably magnified, but stacking of 
multiple light-curves and observations of microlensing on
cosmological scales may be able to discriminate between \gr\ and 
non-\dm\ theories. 
Galaxy-galaxy lensing is likely to be an even more powerful
probe of gravity, with the Sloan Digital Sky Survey (\SDSS)
commissioning data used here to constrain the deflection
law of galaxies to be $A(R) \propto R^{0.1 \pm 0.1}$
for impact parameters in the range  50 kpc $\la R \la$ 1 Mpc.
Together with observations of flat rotation curves,
these results imply that,
in any gravitational theory,
photons must experience (close to) twice the
deflection of massive particles moving at the speed of light
(at least on these physical scales).
The full \SDSS\ data-set will also be sensitive to asymmetry 
in the lensing signal and variation of the deflection law with
galaxy type. A detection of either of these effects 
would represent an independent confirmation that galaxies are
\dm-dominated; conversely azimuthal symmetry of the shear
signal would rule out the typically ellipsoidal haloes predicted 
by most simulations of structure formation.
\end{abstract}


\section{Introduction}
\label{section:intro}

The hypothesis that the universe is made up primarily of luminous,
baryonic matter and that its dynamics are governed by general relativity (\gr) 
is manifestly incorrect.
Measurements of galaxy rotation curves 
provide the clearest invalidation
of the above model, but
the dynamics of galaxy clusters 
and the uniformity of the cosmic microwave
background radiation 
both lead to the same conclusion (as reviewed in \eg\ Trimble 1987
or Peebles 1993).
An obvious possibility is that the universe contains large amounts
of (as-yet undetected) \dm, and this hypothesis has become
part of the standard cosmological model.

The alternative to \dm\ is that \gr\ 
is incorrect on the large scales or low accelerations
not subject to
direct investigation within the Solar system, 
and a number of possible alternative gravity theories
have been proposed (\eg\ 
Milgrom 1983;
Tohline 1983;
Beckenstein \& Milgrom 1984; 
Rood 1984;
Mannheim \& Kazanas 1989;
Beckenstein \& Sanders 1994).
A generic feature of these theories is that the gravitational
force of an isolated \pointmass\ 
is modified to fall off more slowly than Newton's inverse square
law, although 
the force law must eventually return to the Newtonian form (or steeper),
to ensure that the gravitational acceleration caused by an ensemble of
masses is bound (\eg\ Walker 1994).

In the absence of \dm\ the nature of gravity can be 
inferred from dynamical observations or gravitational lensing.
Aside from their tendency
to rely on assumptions of equilibrium, dynamical measurements 
are subject to the more fundamental limitation that the gravitational field 
can only be probed in regions where there is visible matter. 
Conversely, gravitational lensing can be used to measure gravitational
effects well beyond the visible extent of the deflector(s). 
If such measurements can be made sufficiently far from the 
lens, any internal structure can be ignored, and it can be treated 
as a \pointmass. 
Such `simple lensing' scenarios
allow the variation of the deflection angle with impact parameter 
to be measured directly.
Once this function is known a number of other tests are possible:
it should have the same form for all simple 
lenses and it should not depend on the orientation 
of the deflector.
Observations of this nature have the power not only to distinguish
between various alternative gravity theories, but also to 
confirm the existence of \dm\ independently.

Mortlock \& Turner (2001) use some of these ideas to investigate
gravitational lensing in the context of 
modified Newtonian dynamics (\mond; Milgrom 1983), but a more 
general approach is adopted here.
A simple parameterisation of the 
deflection law of a \pointmass\ (\sect{theory})
is constrained using two simple lensing scenarios:
galaxy-galaxy lensing (\sect{galgal})
and microlensing (\sect{micro}).
These results and the future possibilities are summarised in 
\sect{conc}.

\section{Generalised deflection laws}
\label{section:theory}

In time it may be possible to invert the available lensing
data to give the deflection law of a \pointmass\
in a model-independent fashion, but for the moment a parameterisation
is required.
This can specified in terms of the reduced bending angle,
$\vect{\alpha}(\vect{\theta})$,
which relates the angular
position of a source, $\vect{\beta}$, to the angular position of 
its image(s), $\vect{\theta}$, via the lens equation, 
$\vect{\beta} = \vect{\theta} + \vect{\alpha}(\vect{\theta})$
(\eg\ Schneider, Ehlers \& Falco 1992). 
Given that only `simple' deflectors (\ie\ ideally \pointmasses,
but also physically extended lenses, such as galaxies,
if the impact parameter is large) 
are considered here, rotational symmetry can be assumed, 
and so the vector notation can be omitted without loss of generality.

The deflection law for a point-mass in \gr\ is simply 
(\eg\ Schneider \etal\ 1992)
\begin{equation}
\label{equation:alpha_gr}
\alpha(\theta) = - \frac{\theta_{\rmn E}^2}{\theta},
\end{equation}
where $\theta_{\rmn E}$ 
is the Einstein radius of the lens, given by
\begin{equation}
\label{equation:theta_e}
\theta_{\rmn E} = \sqrt{\frac{4 G M}{c^2} \frac{d_{\rmn os}}
{d_{\rmn od} d_{\rmn ds}}},
\end{equation}
where $G$ is Newton's gravitational constant,
$M$ is the mass of the deflector, 
$c$ is the speed of light,
and 
$d_{\rmn od}$, $d_{\rmn os}$ and $d_{\rmn ds}$ are the angular
diameter distances from observer to deflector, observer to source,
and deflector to source, respectively. 
For a given cosmological model these distance measures are
well defined in \gr, but it is unclear how they vary with
redshift in an alternative theory. 
Fortunately the results presented here are not strongly dependent 
on the distance measure used,
and so they are calculated assuming a standard Einstein-de Sitter
model (with Hubble's constant taken to be $H_0 = 70$ km s$^{-1}$ Mpc$^{-1}$).
Nonetheless a more self-consistent formulation should be adopted 
when a full analysis of this sort becomes feasible.

If \gr\ is incorrect on large scales (or at small accelerations)
the deflection law given in \eq{alpha_gr} will also break down, 
presumably falling off more slowly with $\theta$. 
In order to preserve generality no specific alternative gravity
theory is adopted here, and instead
a more generic point-mass deflection law is adopted.
The parameterisation used is 
\begin{equation}
\label{equation:alpha_general}
\alpha(\theta) = - \frac{\theta_{\rmn E}^2}{\theta}
\left(\frac{\theta_0}{\theta_0 + \theta} \right)^{\xi - 1},
\end{equation}
which matches the Schwarzschild form for $\theta \ll \theta_0$,
but falls off as $\alpha(\theta) \propto \theta^{- \xi}$ 
for $\theta \gg \theta_0$.
Note that the deflection angle actually increases with impact
parameter if $\xi < 0$.
In terms of a physical theory, characterised by a 
scale $r_0$ beyond which the physics becomes non-Newtonian, 
\eq{alpha_general}
suggests the identification $\theta_0 = r_0 / d_{\rmn od}$.
\Gr\ (given by $\xi = 1$) is scale-free, and so 
$r_0$ can take any value;
in a non-\dm\ theory
dynamical measurements of galaxies imply 
that $r_0 \simeq 10$ kpc and $\xi \simeq 0$, although
this scale may either vary with mass
[\eg\ in \mond\ $R_0 = (G M / a_0)^{1/2}$, where 
$a_0 \simeq 1.2 \times 10^{-10}$ m s$^{-2}$; Milgrom 1983] 
or be a fundamental constant of the theory.

From the deflection law defined in \eq{alpha_general},
the tangential shear of an image
is given by (\cf\ Miralda-Escud\'{e} \etal\ 1991; Fischer \etal\ 2000)
\begin{eqnarray}
\label{equation:shear}
\gamma_{\rmn tan}(\theta) & \simeq &
\frac{1}{2} \left[ \frac{{\rmn d}\alpha}{{\rmn d}\theta}
-
\frac{\alpha(\theta)}{\theta} \right]
\\
& = & \frac{\theta_{\rmn E}^2}{\theta_0 \theta^2}
\left(
\frac{\xi + 1}{2} \theta + \theta_0
\right)
\left(
\frac{\theta_0}{\theta_0 + \theta}
\right)^{\xi} . \nonumber
\end{eqnarray}
Note that this is half the image polarisation, $p(\theta)$,
as defined by Brainerd \etal\ (1996).
If $\theta \gg \theta_0$ then
\eq{shear} reduces to
$\gamma_{\rmn tan}(\theta) = (\xi + 1) / 2 \,
(\theta_{\rmn E} / \theta_0)^2
(\theta_0 / \theta)^{\xi + 1} \propto \theta^{- (\xi + 1)}$.

The magnification of an image is given by
(\eg\ Sch\-neider \etal\ 1992)
\begin{eqnarray}
\label{equation:mu}
\mu(\theta) & = & \left| \left[1 + \frac{\alpha(\theta)}{\theta} \right]
\left[1 + \frac{{\rmn d}\alpha}{{\rmn d}\theta} \right] \right|^{-1} \\
& = &
\left| \frac{\theta^2 \theta_0 (\theta_0 + \theta)^\xi}
{\theta_{\rmn E}^2 \theta_0^\xi (\theta_0 + \theta)
- \theta^2 \theta_0 (\theta + \theta_0)^\xi} \right| \\
& \times &
\left| \frac{\theta^2 \theta_0 (\theta_0 + \theta)^\xi}
{\theta_{\rmn E}^2 \theta_0^\xi (\xi \theta + \theta_0)
+ \theta^2 \theta_0 (\theta + \theta_0)^\xi} \right|
, \nonumber
\end{eqnarray}
from \eq{alpha_general}.
From this
it is also possible to calculate the total magnification
of a source,
$\mu_{\rmn tot} (\beta)$, 
by solving the lens equation
and then summing the magnifications of the resultant images.

Other observables could be calculated, but the 
shear of an image
and 
the total magnification of a source 
are chosen as they relate directly to the quantities 
observed in measurements of 
galaxy-galaxy lensing (\sect{galgal})
and 
microlensing (\sect{micro}), respectively.

\section{Galaxy-galaxy lensing}
\label{section:galgal}

Background galaxies are observed to be tangentially aligned 
around foreground galaxies due to the latter
population's gravitational lensing effect. 
The angular dependence of the shear signal
is consistent with the hypothesis
that galaxies are dominated by approximately isothermal haloes 
(\eg\ Brainerd \etal\ 1996; Fischer \etal\ 2000), but could
also be explained 
without recourse to \dm\ if 
the (effective)
gravitational force decreases as $r^{-1}$ at large distances.

Under the no-\dm\ hypothesis galaxy-galaxy lensing 
is a very clean probe of the deflection law.
Further, as such measurements rely on averaging over many
background sources, the signal is only appreciable 
at large angular separations from the foreground deflectors. 
Thus, in the absence of \dm, the foreground galaxies can be 
regarded as simple lenses and, despite the fact that \eq{alpha_general}
does not match their lensing properties at small $\theta$,
and the available data can be used to constrain the deflection
law directly.
The most comprehensive galaxy-galaxy lensing observations
made to date are those described by Fischer \etal\ (2000),
who used Sloan Digital Sky Survey (SDSS; York \etal\ 2000) commissioning
data to measure the mean shear signal out to 
$\sim 600$ arcsec around $\sim 3 \times 10^4$ foreground galaxies.
A power law fit of the form 
\begin{equation}
\label{equation:powerlaw}
\gamma_{\rmn tan}(\theta) = \gamma_{\rmn tan,60}
\left(\frac{60 \,\, {\rmn arcsec}}{\theta}\right)^{\eta}
\end{equation}
gave one standard deviation limits of 
$\gamma_{\rmn tan,60} = 0.0027 \pm 0.0005$
and
$\eta = 0.9 \pm 0.1$, although the errors are correlated.

Comparing \eqs{shear} and (\ref{equation:powerlaw}), 
it is clear that $\theta_0$ and $\xi$ are directly
constrained by these results, provided 
that $\theta_{\rmn E}$ is known.
It is given by integrating over the deflector and source
populations (\cf\ Brainerd \etal\ 1996) an approach that will
be necessary when the full \SDSS\ data-set becomes available.
However $\theta_{\rmn E}$ can usefully be
approximated by a fiducial value here as 
the shear signal (and hence the deflection law)
is a simple power law in the regime probed,
and so the angular dependence of \eq{alpha_general} can be factored out
of the integrals, which then only give the normalisation.
Moreover, the range of angular separations is such that the fitted
values of $\theta_0$ and $\theta_{\rmn E}$ are degenerate, and so 
the normalisation of the signal cannot place strong limits on 
the average mass of the foreground galaxies.
The local galaxy population is dominated by spirals
(\eg\ Postman \& Geller 1984) and the mean deflector and source
redshifts are $\langle z_{\rmn d} \rangle = 0.17$
and $\langle z_{\rmn s} \rangle = 0.3$, respectively (Fischer \etal\ 2000),
which imply that $\theta_{\rmn E} = 1.0 \pm 0.1$ 
arcsec\footnote{The value is slightly larger than the actual
Einstein radii of the galaxies due to use of the \pointmass\ 
lens model, and the large error is a combination of
uncertainties
in the distance measures and the size of the foreground galaxies.}.

\begin{figure}
\includegraphics{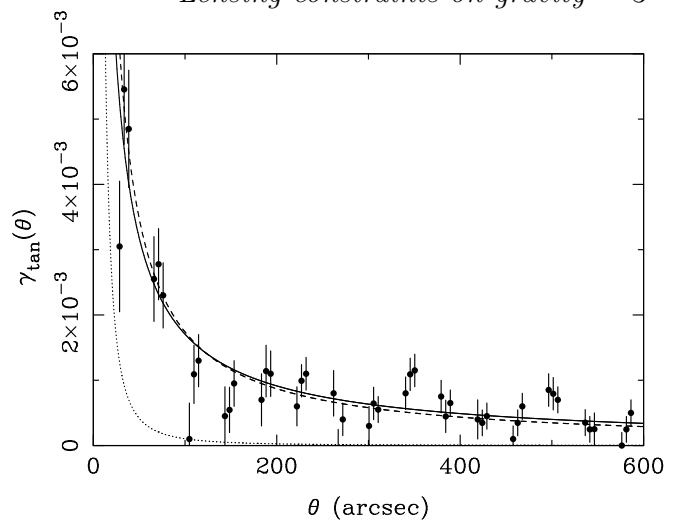}
\vspace{\singlefigureheight}
\caption{The mean shear around foreground galaxies
in the $g^\prime$, $r^\prime$ and $i^\prime$ bands,
as measured by Fischer \etal\ (2000)
compared with various theoretical
predictions.
The data in the three bands are offset for clarity,
and the the three models 
(all of which assume $\theta_{\rmn E} = 1$ arcsec)
are:
$\xi = - 0.1$ and $\theta_0 = 3.7$ arcsec
(the best-fit model; solid line);
$\xi = 0$ and $\theta_0 = 2.9$ arcsec
(\mond\ or isothermal \dm\ haloes; dashed line);
and
$\xi = 1$
(\gr\ with no \dm, which is independent of $\theta_0$; dotted line).}
\label{figure:galgalresults}
\end{figure}

Using the above value for $\theta_{\rmn E}$,
\eq{powerlaw} implies the constraints
$\theta_0 = 3.65 \pm 0.08$ arcsec
(which implies that $r_0 = d_{\rmn od} \theta_0
= 10 \pm 2$ kpc, assuming $d_{\rmn od} = 600$ Mpc)
and $\xi = - 0.1 \pm 0.1$, 
results confirmed by an independent likelihood analysis.
Several of these fits, along with the Fischer \etal\ (2000)
data, are shown in \fig{galgalresults}.
Firstly, it is important to note that the data points
cover angles much greater than the inferred value of $\theta_0$;
if this were not the case \eq{alpha_general} could not be used 
for the deflection law.
As expected, \gr\ (\ie\ $\xi = 1$)
cannot explain the signal without recourse 
to \dm, but the data are consistent with the 
\mond ian lensing formalism investigated by Qin \etal\ (1995)
and Mortlock \& Turner (2001),
which predicts $\xi = 0$. 
More generally, the concordance between dynamical measurements
and these lensing results implies 
that the relativistic prediction for geodesics 
(that photons experience twice 
the deflection of massive particles moving at the speed of light)
must also be true in any alternative theory of gravity.

The observations used in the above analysis represent just 
a few percent of the eventual \SDSS\ data-set, 
which should allow shear measurements out to several Mpc
with uncertainties at about the 5 per cent level.
Beyond this scale the signal will be diluted by secondary
deflectors along the \los, although if the shear signal 
was observed to drop off faster than 
$\gamma_{\rmn tan}(\theta) \propto \theta^{-1}$ it could indicate
the edge of galaxy haloes (in the \dm\ paradigm)
or
a return Newtonian physics (if there is no \dm). 

The \SDSS\ data will also allow 
the azimuthal symmetry of the mean shear signal 
(\cf\ Natarajan \& Refregier 2000)
to be measured; this provides a means of distinguishing between
the two paradigms.
A rotationally invariant signal would imply that
\dm\ haloes are typically spherical (or at least circular in projection),
in conflict with most collapse models
(\eg\ Navarro, Frenk \& White 1995, 1996),
whereas
any measured
asymmetry 
much beyond the visible extent of the foreground galaxies
would be difficult to reconcile with the hypothesis they contain
no \dm. 

Further, non-\dm\ theories,
by virtue of their simplicity, 
are also subject to a number of tests that have no counterpart if
there is \dm.
Modulo scaling uncertainties (see \sect{theory}),
the \pointmass\ deflection law should always have the same form.
In the context of galaxy-galaxy lensing this means that 
the shear profile should be the same, within statistical uncertainties,
around all possible subsets of the foreground galaxy population. 
For instance, whilst spirals and ellipticals may have slightly different
mass-to-light ratios, their deflection laws should have 
essentially the same form.
This principle can also be extended to stellar mass lenses, a
possibility explored in \sect{micro}.

\section{Microlensing}
\label{section:micro}

Microlensing of a point-source by a single star provides 
an unambiguous measurement of the deflection law of a \pointmass,
but does not necessarily probe regimes in which gravity could
be expected to deviate from \gr.
The family of theories described in \sect{theory} become
non-Newtonian only beyond some scale $r_0$, which must be 
several kpc for galaxies but may or may not vary with the
mass of 
the deflector. If it does not, then an isolated star would 
behave as a Schwarzschild lens out to several kpc, 
whereas microlensing measurements are only likely to probe
sub-pc impact parameters.
However, if $r_0$ scales with mass, 
the deflection law of stars could differ from the standard 
form as close in as $\sim 0.01$ pc (\eg, $r_0 \simeq 0.03$ pc
for a Solar mass star in \mond). 
As shown in 
Mortlock \& Turner (2001),
microlensing light-curves in such a theory would have 
narrower peaks and broader wings (relative to \gr), 
although if this was observed, care would have to be taken
to exclude more mundane explanations, such as blending and finite
source effects.

Observationally, most efforts to detect microlensing have been
concentrated in the local group.
Several collaborations 
(\eg\ Afonso \etal\ 1999; Alcock \etal\ 2000; see Paczynski 1996 for a
summary) 
have monitored millions of stars in the Magellanic Clouds 
and the Galactic bulge,
and over a hundred events have been recorded.
In several cases the deflector appears to be a binary 
system (\eg\ Alcock \etal\ 2001), but for 
the vast majority the approximation that
the lens is a single, isolated \pointmass\ is excellent. 
Unfortunately, the distance scales within the local 
group are such that single light-curves can only
probe Solar system scales, a regime in which Newtonian physics 
has already been confirmed. 
Even if magnifications of 0.1 per cent could be measured,
microlensing of sources in the Large Magellanic Cloud 
(\LMC; $d_{\rmn os} \simeq 60$ kpc) would still 
only probe the gravitational field of a Solar mass 
deflector to scales of $\sim 0.001$ pc.
One way to escape this limitation might be to `stack' the 
light-curves of a large number of lens events, although 
the effective integration over the deflector population could
dilute any non-Newtonian signatures.

A more direct way around this geometrical problem is to 
search for low optical depth microlensing at cosmological distances.
From the definitions in \sect{theory}, 
and assuming the deflector to be about half way between observer and source,
$\theta_{\rmn E} \propto d_{\rmn od}^{- 1/2}$, whereas 
$\theta_{\rmn 0} \propto d_{\rmn od}^{-1}$, which
together imply that 
$\theta_{\rmn 0} / \theta_{\rmn E} \propto d_{\rmn od}^{-1/2}$. 
Thus at cosmological distances
$\theta_{\rmn 0} \la \theta_{\rmn E}$ and the strong lensing
regime is subject to non-Newtonian effects, resulting in 
microlensing light-curves which are markedly more peaked, as described above. 
The degree of distortion depends upon the value of $r_0$, 
but any simple microlensing
event with a source redshift close to unity should differ visibly
from the \gr\ prediction if $r_0 \la 0.1$ pc (Mortlock \& Turner 2001).

It is possible that microlensing of a
cosmologically distant source has already been seen,
albeit serendipitously.
The redshift 2.04 gamma ray burst (\grb) 000301C (Sagar \etal\ 2000)
was observed to have
an achromatic peak in its otherwise smoothly decaying
light-curve. Garnavich \etal\ (2000) successfully modelled this
as microlensing of the expanding fireball by a \pointmass, although
the fit relied on the assumption that it appeared as a ring on the sky.
Unfortunately the photometry of \grb\ 000301C was not of sufficient
quality to facilitate a measurement of the deflection law of the lens,
and the uncertain nature of the source only makes such inferences
more difficult.
More lensed \grbs\ should be discovered 
(even if the event rate is low; Koopmans \& Wambsganss 2001),
but there are also systematic cosmological microlensing searches underway.

Both Walker (1999) and Tadros, Warren \& Hewett (2001) describe
programs to monitor high-redshift quasars seen through 
the outskirts of nearby galaxies and clusters. 
There have not yet been any detection of microlensing, 
but even a single light-curve should be sufficient to place
a lower bound on $r_0$ (assuming $\xi \simeq 1$; see \sect{theory}). 
However the motivation for both these searches was to
search for compact \dm\ in the foreground objects
and so some of the sources were chosen to lie behind their `outer haloes',
regions which might be completely devoid of microlenses if there is no \dm. 
This would be unfortunate in the context of light-curve
measurements, but also suggests an alternative test of non-\dm\
theories.

If there is no \dm, then the only potential deflectors on
these scales are planets, stars and other (known) compact objects.
Observations of `dark' microlensing would tend contradict this 
hypothesis, but no such detection has been reported.
The local group results are consistent with there being do compact
\dm\ in the galactic halo, provided that the \LMC\
is extended along the \los\ (Afonso \etal\ 1999; Alcock \etal\ 2001).
The quasar monitoring programs described above have not yet 
detected any lensing at all, and the \grb\ microlens could be  
a star in a galaxy near the \los\ (Garnavich \etal\ 2000). 
Microlensing has been observed in Q~2237+0305
(\O stensen \etal\ 1996; Wo\'{z}niak \etal\ 2000), but this 
can be attributed to stars in the bulge of the lensing galaxy,
and no microlensing has been observed in either image of
Q~0957+561 (\eg\ Pelt \etal\ 1998).
These results are not only consistent with the no-\dm\ hypothesis,
but also strongly rule out several popular \dm\ candidates.

\section{Conclusions}
\label{section:conc}

If there is no \dm\ in the universe 
then gravitational lensing is an ideal probe 
of the gravitational field around isolated deflectors. 
By applying basic symmetries, lensing can be used to distinguish between
alternative gravity theories and \gr,
or used to measure the deflection law in non-\dm\ models.

The most suitable data available are the galaxy-galaxy lensing 
measurements of Fischer \etal\ (2000). 
Modelling the deflection law of the foreground galaxies as a power
law, the logarithmic slope was constrained to be $0.1 \pm 0.1$
beyond $\sim 50$ kpc.
The full \SDSS\ data-set will be fifty 
times larger, facilitating a measurement of the (as)symmetry of
the shear signal. This is particularly powerful diagnostic,
with the power to discriminate unambiguously between the 
\dm\ paradigm and alternative gravity theories.

Non-\dm\ models are subject to several other tests, based on 
the principle that the deflection laws of all isolated deflectors
should be the same, modulo possible scaling uncertainties. 
The galaxy-galaxy lensing results should be independent of galaxy
type or luminosity, and the implied physics should be
consistent with that inferred from lensing by stars and planets.
Observations of microlensing within the local group are 
fairly insensitive to any putative departures from \gr, as 
non-Newtonian effects are expected only in the wings of lensing
events. Nonetheless it may be possible to synthesise
sufficiently accurate
photometry by stacking multiple light-curves. 
More promising is the prospect of measuring microlensing 
events on cosmological scales, either by serendipitous discoveries
(Garnavich \etal\ 2000) or dedicated monitoring programs
(Walker 1999; Tadros \etal\ 2001). 
With good photometry even a single light-curve could be used to 
measure the deflection law of a star out to many pc, scales 
on which many theories predict non-Newtonian effects.

All the above ideas pertain to simple lensing scenarios,
in which the deflector
is not spatially extended; more complex situations 
permit various observational tests as well, but require additional
theoretical development (\cf\ Mortlock \& Turner 2001). 
Moreover the tests proposed above are powerful primarily due 
to their simplicity, and the forthcoming results should 
place unambiguous constraints on the nature of gravity.

\section*{Acknowledgments}
The authors acknowledge a number of interesting discussions with
Anthony Challinor,
Geraint Lewis,
Bohdan Paczy\'{n}ski
and 
Joachim Wambsganss.
Several aspects of this paper were clarified due to the comments of
the anonymous referee.
DJM was funded by PPARC
and 
this work was supported in part by NSF grant AST98-02802.

\bsp
\label{lastpage}
\end{document}